\documentclass[a4paper]{jpconf}
\usepackage{amsmath}
\usepackage{graphicx}
\begin{document}

\title{Towards a Predictive HFB+QRPA Framework for Deformed Nuclei: Selected Tools and Techniques}

\author{Emanuel V. Chimanski $^{1}$, Eun Jin In$^{1}$, Jutta E. Escher $^{1}$, Sophie P{\'e}ru$^{2}$ and Walid Younes$^{1}$}

\address{Lawrence Livermore National Laboratory, Livermore, CA, USA $^{1}$ \\
 CEA, DAM, DIF, Arpajon, France $^{2}$}

\ead{chimanski1@llnl.gov}

\begin{abstract}
Reliable predictions of the static and dynamic properties of a nucleus require a fully microscopic description of both ground and excited states of this complicated many-body quantum system. Predictive calculations are key to understanding such systems and are important ingredients for simulating stellar environments and for enabling a variety of contemporary nuclear applications. Challenges that theory has to address include accounting for nuclear deformation and the ability to describe medium-mass and heavy nuclei.  Here, we perform a study of nuclear states in an Hartree-Fock-Bogoliubov (HFB) and Quasiparticle Random Phase Approximation (QRPA) framework that utilizes an axially-symmetric deformed basis. We present some useful techniques for testing the consistency of such calculations and for interpreting the results.
\end{abstract}

\section{Introduction}

Achieving predictive descriptions of the structure and dynamics of atomic nuclei is a challenge that has a long history of progress and insights, but that has not lost relevance.  New experimental facilities, such as radioactive beam facilities, are poised to provide us with data on exotic isotopes that require interpretation~\cite{Balantekin-2014,Johnson-2020,Hilaire2021}, and multi-physics simulations of extreme environments, such as supernovae and neutron-star mergers need information on nuclear properties.  The quest to understand the generation of energy in stars, stellar evolution, and the production of the elements relies on nuclear masses, nuclear decay properties, and nuclear reaction rates~\cite{Arcones:17}.  

Nuclear theory comprises a wide range of description, from phenomenological, such as the liquid drop model~\cite{Mottelson-97}, to sophisticated microscopic approaches~\cite{Hergert:20,Ring-2004,Suhonen-2017}.  An important goal is to establish microscopic descriptions that are able to predict ground-state properties, low-energy excitations, and high-energy collective modes for a broad range of nuclei.  Ideally, the underlying  interactions connect to QCD or utilize a small number of parameters that are fitted to a limited set of nuclear properties.  The latter approach is employed by density-functional theories and the quasi-particle random-phase approximation (QRPA), which is the focus of this contribution~\cite{Ring-2004,Suhonen-2017,Younes-2019}.  Specifically, we are using the finite-range Gogny interaction to establish a consistent description of both spherical and deformed nuclei across a wide range of the isotopic chart.  Our calculations are performed in an axially-symmetric deformed harmonic-oscillator basis. The formalism and implementation is inspired by and tested against the ground-breaking development published in Refs.~\cite{Peru-2014,Deloncle-2017,Peru-2019,Peru-2021}.

The purpose of the current contribution is to establish some tools to test and better interpret the calculations.  In particular, we discuss angular-momentum restoration from the deformed calculations, illustrate that the calculations exhibit the proper behavior in the spherical limit, and compare some results to the well-established RPA approach in the spherical limit for the closed-shell nucleus $^{16}$O.

\section{Formalism}
\label{sec:formalism}
Our theoretical framework is the QRPA in an axially-symmetric deformed harmonic-oscillator basis that allows for the use of the finite-range Gogny interaction \cite{Goriely-2009}. The QRPA excitations are built on top of Hartree-Fock-Bogoliubov (HFB) description of the nuclear ground state.  Both HFB and QRPA make consistent use of the same interaction. We restore angular-momentum symmetry in order to study transitions from ground to various excited states.  We also investigate the spatial behavior of the static and transition densities.

\subsection{QRPA approach in a deformed basis}
\label{subsec:qrpa}

The well-known Bogoliubov transformation replaces bare particles, represented by Latin symbols, by quasi-particle (qp) states, represented by Greek symbols. 
This transformation can be inverted to give the occupation of single particles states ($\alpha$) in terms of linear combinations of creation  and annihilation of quasi-particle ($i$) modes 
\begin{eqnarray}
\label{sp_qp}
c_{\alpha}^{\dagger}= \sum _{i}\left [ U^{*}_{\alpha i} \eta ^{\dagger}_{i}+V_{\alpha i} \eta _{i} \right ], \quad c_{\alpha}= \sum _{i}\left [ V^{*}_{\alpha i} \eta ^{\dagger}_{i}+U_{\alpha i} \eta _{i}\right ]. \end{eqnarray}
By definition, the HFB vacuum is destroyed by the application of the quasiparticle annihilation operator: $\eta _{i} |\text{HFB } \rangle =0$.

The HFB density matrix and pairing tensor are given in terms of the coefficients $U$ and $V$ as
\begin{equation}
\rho _{\beta \alpha}=\sum _{i}V_{\alpha i}V^{*}_{\beta i} \qquad \kappa _{\beta \alpha}=\sum _{i}U_{\beta i}V^{*}_{\alpha i}  
\label{densities}.
\end{equation}
 The HFB mean-field is obtained variationally, with constraints on particle number and quadrupole deformation
\begin{equation}
    \beta = \sqrt{\frac{5}{9}\pi}\frac{q_{20}}{AR^{2}},
\end{equation}
where $R=1.2 A^{\frac{1}{3}}$ is the nuclear radius,  $A$ the number of nucleons and $q_{20}$ the mean value of the axial quadrupole operator. All other parameters are kept fixed.
Further details about the HFB theory, matrix elements, and the numerical algorithm employed here can be found in Refs.~\cite{Ring-2004,Younes-2009,Younes-2019}.
 
We focus on nuclear excitations that can be represented by in the QRPA framework. The QRPA excited states are generated through the action of the creation operator
\begin{eqnarray}
\hat{\theta }^{\dagger}_{n,K} = \sum _{i<j}\left [ X^{ij}_{n,K}\eta ^{\dagger}_{k_{i}}\eta ^{\dagger}_{k_{j}} + Y^{ij}_{n,K}\eta _{-k_{j}}\eta _{-k_{i}} \right ]
  \end{eqnarray}
on the QRPA vacuum state:  $|\theta_{n},K \rangle = \hat{\theta }^{\dagger}_{n,K}|0_{\text{def}},(K=0)\rangle $, where $\theta _{n}|0_{\text{def}},(K=0)\rangle = 0$.
The $X$ and $Y$ coefficient matrices are obtained by solving the well-known QRPA matrix equation
 \begin{equation}
 \label{QRPA_matrix}
\begin{bmatrix}
A           & B  \\
B ^{*} & A^{*}
\end{bmatrix}
\begin{bmatrix}
X \\
Y
\end{bmatrix}
=
\omega 
\begin{bmatrix}
I & 0 \\
0 &-I
\end{bmatrix}
\begin{bmatrix}
X \\
Y
\end{bmatrix},
 \end{equation}
in a basis of 2 quasi-particle (2qp) states for a given projection of angular momentum and parity $K^{\pi}$ with $K = k_{i} +k_{j}$ and $\pi = \pi _{i}\,\pi _{j}$. The details about the QRPA matrix elements expressions can be found in~\cite{Peru-2014,Ring-2004,Suhonen-2017}. The 2qp states are generated consistently by using the same Gogny interaction that is employed at the HFB level, here D1M. 

We restrict our applications to even-even nuclei, and consider $K=0$ vacuum states only.  Consequently, the QRPA excitations have well-defined $K$ quantum numbers, but not good angular momentum.

\subsection{Angular Momentum Restoration}
\label{subsec:restoration}
States of good angular momentum $|JM(K)_{n} \rangle$ can be constructed from the intrinsically excited states $|\theta_{n},K \rangle$ by using the Wigner rotation matrices.  To obtain the response to a transition operator with well-defined angular momentum and parity $J^{\pi}$, one needs to calculate QRPA excitations $K^{\pi}=0 ^{\pi},\pm 1^{\pi},\ldots, \pm J^{\pi}$ and use angular-momentum restoration to determine their contributions~\cite{Ring-2004,Peru-2014}:
\begin{eqnarray}
\label{projected_response}
  \langle \tilde{O}_{(J^{\pi}=0^{+})} |\hat{Q}_{\lambda \mu}|JM(K)_{n} \rangle =  \hat{J}  \begin{pmatrix}
  0  & \lambda & J\\
  0  &  -\mu & M 
  \end{pmatrix}
  \sum _{\mu ^{\prime}}(-1)^{\mu^{\prime} -\mu }\bigg [ 
\begin{pmatrix}
 0 & \lambda &J \\
 0 & -\mu ^{\prime}& K 
  \end{pmatrix}
  &\langle 0_{\text{def}}|r^{\lambda }Y_{\lambda \mu^\prime}|\theta _{n},K \rangle &\nonumber \\
    +(-1)^{J-K}
  \begin{pmatrix}
 0 & \lambda &J \\
 0 & -\mu ^{\prime}& - K 
  \end{pmatrix}
\langle 0_{\text{def}}| r^{\lambda }Y_{\lambda \mu^\prime}|\theta _{\bar{n}},-K \rangle \bigg ],
\end{eqnarray}
where $\hat{J}\equiv \sqrt{2J+1}$ and $|\theta ^{\dagger}_{\bar{n}},-K \rangle$ represents time-reversed states. The 3-j symbols restricts $J=\lambda$ and $K= \mu ^{\prime}$. For spherical nuclei ($\beta =0$), the different $K^{\pi}$ components are degenerated and it suffices to calculate QRPA $K^{\pi} = 0 ^{\pm}$ states. The degeneracy provides a stringent condition that allow us to test the calculations in the spherical limit, as we will demonstrate for $^{16}$O in this work.

\subsection{Transition densities}
\label{subsec:restoration}
The intrinsic density for a transition from the ground state to a QRPA excited state is given by
\begin{equation}
  \label{td_intrinsic}
\rho ^{n,K}(\vec{r})= \sum _{\alpha \beta } \phi ^{*}_{\alpha}(\vec{r} ) \phi _{\beta}(\vec{r}) \, Z^{n,K}_{\alpha,\beta }.
\end{equation}
where we have defined the spectroscopic amplitude
\begin{equation}
  \label{ZabnK}
Z^{n,K}_{\alpha,\beta }\equiv  \langle \tilde{0}|c^{\dagger}_{\alpha }c_{\beta} |\hat{\theta } _{n,K}\rangle .
\end{equation}
The matrix elements ($\ref{ZabnK}$) can be evaluated to give
\begin{eqnarray}
Z^{n,K}_{\alpha,\beta } =
  &&  \, \sum _{i<j} \bigg [ X^{i,j}_{n,K} \left ( V_{\alpha j}U_{\beta i} -  V_{\alpha i}U_{\beta j}\right  ) + Y^{i,j}_{n,K} \left ( U^{*}_{\alpha i} V_{\beta j} - U^{*}_{\alpha j} V_{\beta i}  \right ) \bigg ].
\end{eqnarray}
Angular-momentum projected transition densities can also be obtained using (\ref{projected_response}). These could be used, e.g., in scattering and reaction calculations.

\section{QRPA dipole response for $^{16}$O}
\begin{figure}[h!]
\centering
  \includegraphics[scale=0.7]{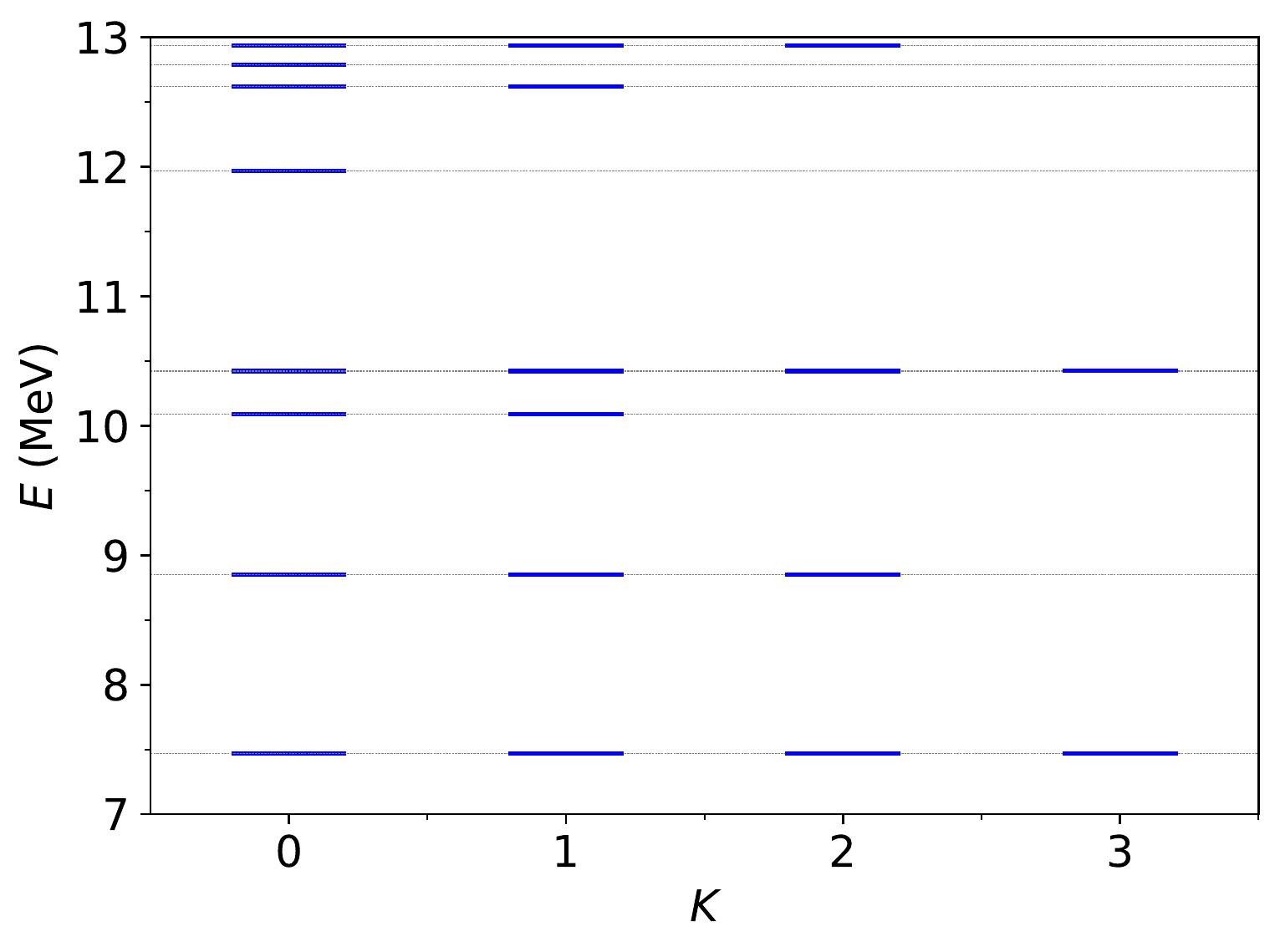}
\caption{\label{Ens}Low energy QRPA spectra for negative parity $K=0^{-},1^{-},2^{-},3^{-}$ states of $^{16}$O. The exact degeneracy of energy values with differing $K$ is a stringent test for calculations in the deformed basis.}
\end{figure}

To test the computation implementation, we carried out calculations for the spherical $^{16}$O nucleus. 
We included $N_{\rm {osc}}=6$ major oscillator shells and 2qp states up to 60 MeV of energy. We performed tests with larger basis sizes with no significant change in the numerical results. Figure~\ref{Ens} shows the low energy part of the QRPA spectrum of $^{16}$O for negative-parity states. As expected, there are degenerate states that belong to different $K$,  but the same $J$ value. This useful consistency test was performed for both positive (not shown) and negative parity (Fig.~\ref{Ens}) levels.

\begin{figure}[h!]
\centering
  \includegraphics[width=6.5cm,height=11cm]{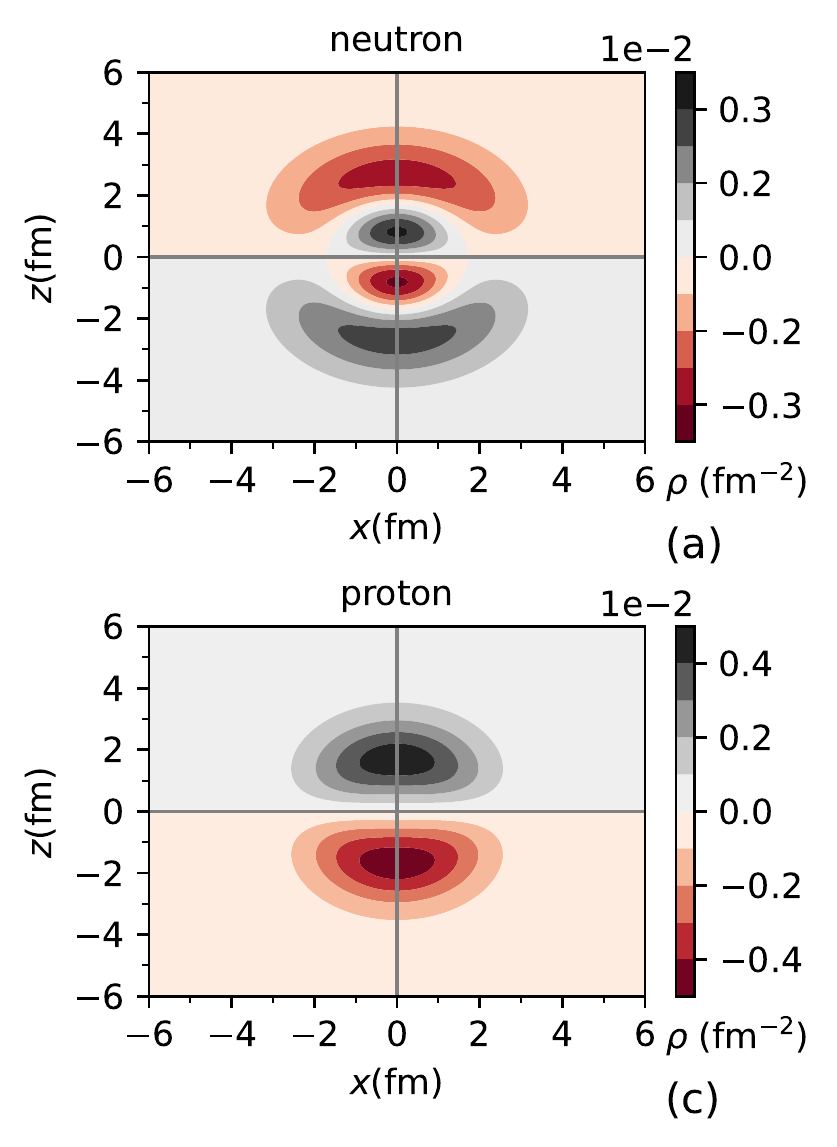}
  \includegraphics[width=6.5cm,height=11cm]{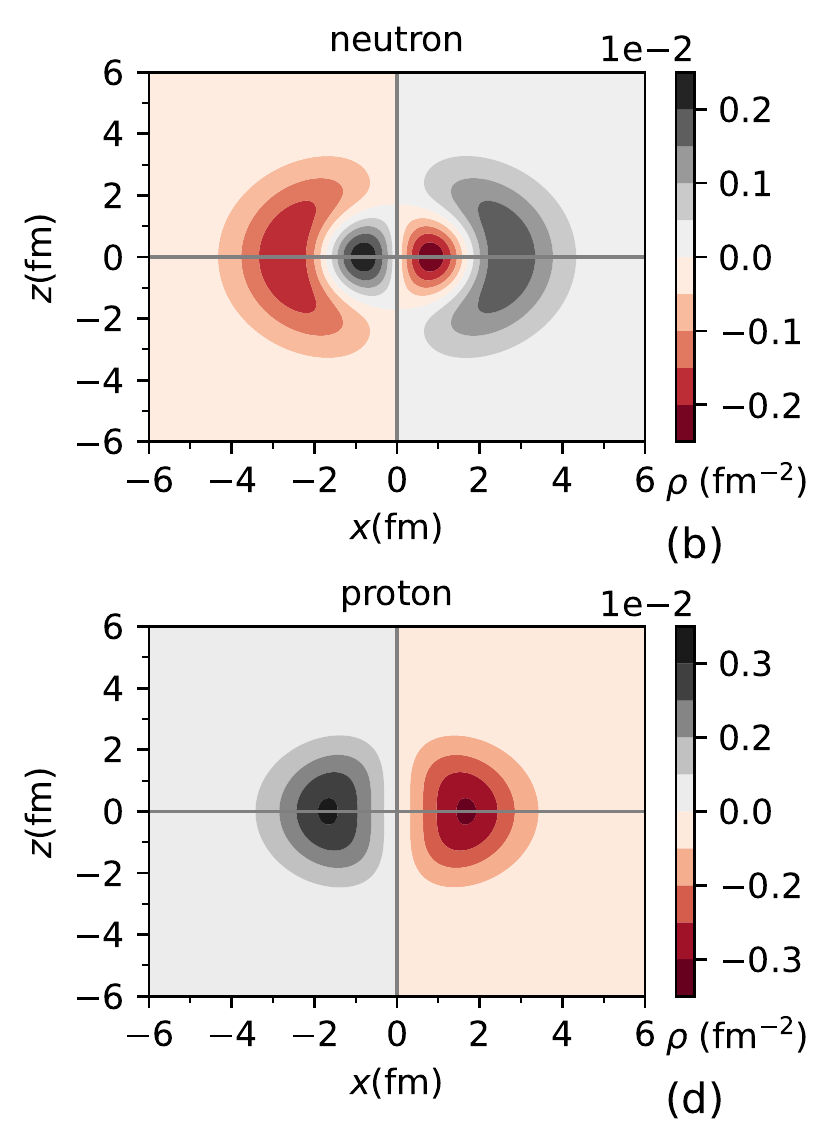}
\caption{\label{intrinsic_td} QRPA intrinsic transition densities for the state at $E=21.65$ MeV for $K=0^{-}$ (a,c) and $K=1^{-}$ (b,d). The state contributes strongly to the dipole response, and the transition density illustrates this character.}
\end{figure}

\begin{figure}[h!]
\centering
  \includegraphics[scale=.9]{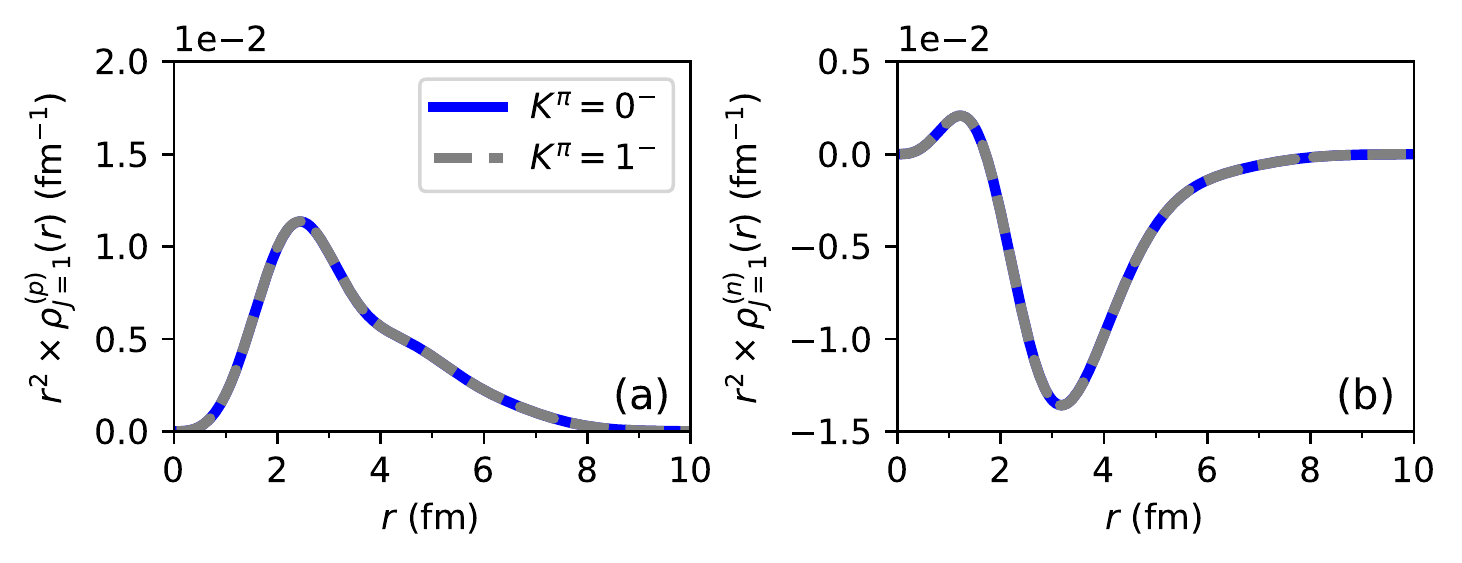}
\caption{\label{GDR_dens_J=1} Projected radial transition densities for the $E=21.65$ MeV state of $^{16}$O.  Panels (a) and (b) show $J=1$ transitions obtained from $K=0^{-}$ (solid blue lines) and $K=1^{-}$ calculations (dashed gray lines) for protons and neutrons, respectively. The agreement of the results from the different $K$ provide a stringent test for the calculations in the deformed basis.}
\end{figure}

\begin{figure}[h!]
\centering
  \includegraphics[scale=0.9]{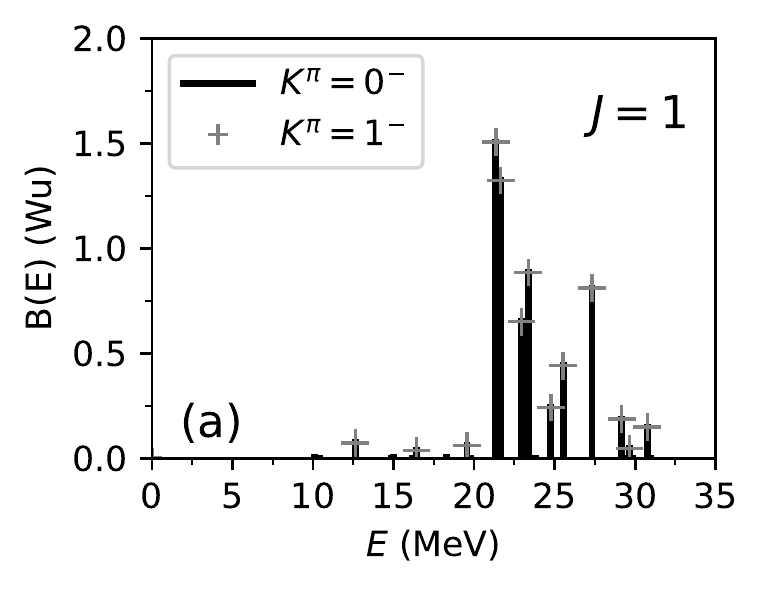}
    \includegraphics[scale=0.9]{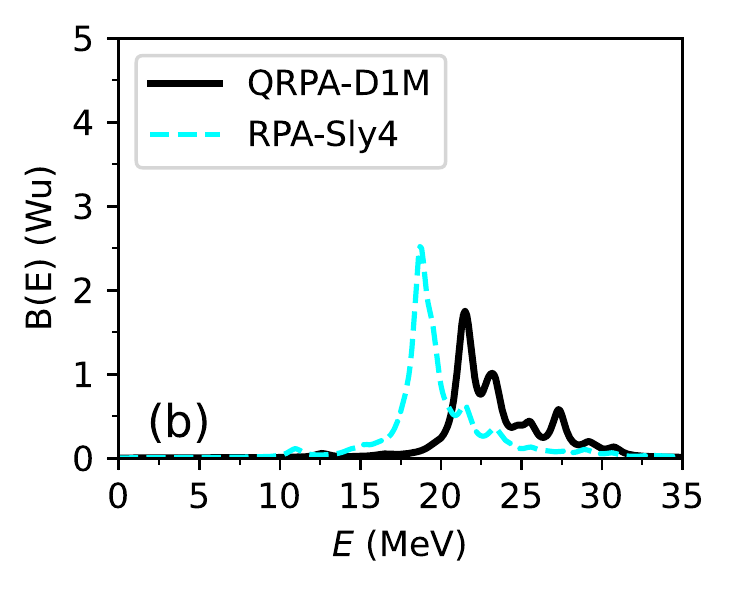}
\caption{\label{16O_IV_J=1}(a) Isovector dipole response for $^{16}$O obtained for $K=0^{-}$ and $K=1^{-}$ QRPA calculations, respectively.  The agreement of the results from the different $K$ calculations provide a test for the calculations in the deformed basis. (b) QRPA-D1M and RPA-Sly4 Isovector dipole responses for $^{16}$O. Transitions are spread with a Breit-Wigner distribution with $\Gamma$ = 1 MeV. The more energetic QRPA response provides a larger EWS value in a better agreement to the systematic expectation (see Table \ref{tab:EWSR}).}
\end{figure}

In Fig.~\ref{intrinsic_td} we show the intrinsic transition densities for the $E=21.65$ MeV state, which contributes strongly to the GDR (Giant Dipole Resonance). Neutron and proton densities were obtained from both $K=0^{-}$ and $K=1^{-}$ QRPA calculations. As expected, the two calculations yield identical densities, apart from an exchange of the $x$ and $z$ axes. These two transitions correspond to the same $J=1$ radial response when angular momentum is restored, as shown in Fig.~\ref{GDR_dens_J=1}.
The perfect matching of the radial shapes from calculations with different $K$ values is an important test for our nuclear structure, transition density and angular momentum restoration implementations.
One can also verify that every degenerate state will produce the same projected response. Figure~\ref{16O_IV_J=1}~(a), shows the response to the dipole operator for QRPA states up to $35$ MeV. There is a perfect agreement between $K=0^{-}$ and $K=1^{-}$ transition strengths. The main peak of the GDR distribution is located at $E=21.65$ MeV.  

Before closing, we present a comparison of our Gogny-QRPA approach with the spherical Skyrme-RPA implementation by Col{\`o} et al. \cite{Colo-2013}. As pairing effects are negligible for the doubly-closed shell nucleus $^{16}$O, differences in the predicted structure are primarily due to the choice of the nuclear interaction. Figure~\ref{16O_IV_J=1}~(b) shows the dipole response for both Gogny-D1M and for Skyrme-Sly4 approaches, the transition strengths were spread with a Breit-Wigner distribution using $\Gamma$ = 1 MeV. The response obtained with Skyrme-Sly4 peaks at slightly lower energies, by a few MeVs, when compared to the Gogny-D1M results. This difference is clear when we compare the Energy-Weighted-Sum-Rule (EWSR) for both approaches. Table~\ref{tab:EWSR} shows both theoretical results and the systematic \cite{Harakeh-2001} expectation for the mean energy values for the GDR. The D1M approach provides $23.4$ MeV and is only about 2 MeV away from the value of $25.4$ MeV expected from the systematics while the Sly4 interaction gives a mean value of 18.4 MeV. Including higher-order effects, i.e. 2p2h or 4-quasiparticle excitations, respectively, is expected to account for some of the spreading that we included phenomenologically and to also shift the position of the peaks.

\begin{table}[h!]
\caption{\label{tab:EWSR}Theoretical and systematic \cite{Harakeh-2001} mean energy values of the isovector giant dipole resonance (GDR) for $^{16}$O. The RPA result is obtained with the RPA-Skyrme code by Col{\`o} et al. \cite{Colo-2013}.}
\centering
\begin{tabular}[t]{lccc} 
\hline
 &QRPA-D1M & RPA-Sly4 & Systematic\\
  \hline
GDR (MEV)& 23.4 & 18.4  &25.4\\
 \hline
\end{tabular}
\end{table}

\section{Outlook}
The present contribution provides useful tests and techniques for performing QRPA calculations with the Gogny interaction in a deformed basis. The degeneracy of energy levels of a spherical nucleus present in this approach was used to verify that our QRPA calculations are correct. We have presented a short review on the intrinsic and radial projected transition densities, quantities that can be employed in reaction calculations.
In addition, we have compared dipole response predictions from Gogny and Skyrme interactions for $^{16}$O. The Gogny D1M interaction produces a GDR mean energy distribution closer to the expected value given by systematic analysis, in contrast to results obtained with the Skyrme-Sly4 parameterization.
Work is underway to use the full approach to study the structure of both spherical and deformed nuclei and make predictions for unstable isotopes as demostrated, e.g, in Refs~\cite{Nobre:10,Nobre:11,Dupuis-2019}.

 \section*{Acknowledgements}
This work is performed in part under the auspices of the U.S. Department of Energy by Lawrence Livermore National Laboratory under Contract DE-AC52-07NA27344 with support from LDRD 19-ERD-017 and LDRD 20-ERD-030.
 
\section*{References}

\bibliographystyle{iopart-num}
\bibliography{lib.bib}

\end{document}